\documentclass{PoS}
\usepackage{graphicx}
\usepackage{amsmath}
\title{Multi-Messenger Searches in Astrophysics}

\ShortTitle{Multi-Messenger Searches in Astrophysics}

\author{\speaker{Kathrin Egberts}\\
        Universit\"at Potsdam, Institut f\"ur Physik und Astronomie, Campus
Golm, Haus 28, Karl-Liebknecht-Str. 24/25, 14476 Potsdam-Golm, Germany\\
        E-mail: \email{kathrin.egberts@uni-potsdam.de}}

\abstract{Multi-messenger astronomy has experienced an explosive development in the past few years. While not being a particularly young field, 
it has recently attracted a lot of attention by several major discoveries and unprecedented observation campaigns covering the entity of the electromagnetic
spectrum as well as observations of cosmic rays, neutrinos, and gravitational waves.
The exploration of synergies is in full steam and requires close cooperation between different instruments. Here I give an overview over 
the subject of multi-messenger astronomy and its virtues compared to classical ``single messenger'' observations, present the 
recent break throughs of the field, and discuss some of its organisational and technical challenges.}

\FullConference{XXIX International Symposium on Lepton Photon Interactions at High Energies - LeptonPhoton2019\\
		August 5-10, 2019\\
		Toronto, Canada}

\begin{document}

\section{Introduction}
Multi-messenger astronomy has experienced an explosive development in the past few years. It is based on the coordinated observation of astrophysical objects in several ``messengers''. These messengers are electromagnetic radiation, cosmic rays, neutrinos, and gravitational waves. They are partly closely interconnected and at the same time provide complementary information on the emitting object, such that  their combination allows for deeper insight into its physical processes than possible with the individual measurements.\\ 
The oldest messenger we have is the photon. Classical astronomy has used observations of the visible sky for centuries to study celestial objects. In the optical, most emission is caused by thermal processes of hot objects like stars, resulting in some black-body 
radiation and the reprocessing of this emission by absorption and re-emission processes. The energy range accessible by photons, however, spans much further than the optical, going from low-energy radio observations of less than meV to
$\sim 100$~TeV $\gamma$-rays, some 15 orders of magnitude in energy. Today the broad-band spectral energy distribution of electromagnetic radiation is measured by a multitude of different instruments, using different techniques both ground-based and in space (for a review see ~\cite{Neronov}). The fact that Nature provides us with such a wide spread of different wavelengths is owed to
the abundance of mechanisms that cause electromagnetic radiation, going beyond the thermal processes mentioned before. {\it Multi-wavelength astronomy} is a long-established field of astrophysics that uses this information to investigate
objects from various perspectives, covering the different mechanisms in place to emit this radiation.
Contrary to the black-body radiation of thermal emission, non-thermal emission processes cover a significantly larger range of energies. 
Their emission originates from the 
interactions of cosmic rays, which are ubiquitous in the Universe. Accelerated in moving magnetic fields, their interaction with matter of interstellar gas and radiation fields produces secondaries, which can be detected themselves: 
\begin{align*}
p + \gamma / p  \longrightarrow  p+\gamma / p +  \pi^+ + \pi^- + \pi^0 
\end{align*}
where the pions can further decay to produce secondary electrons and electron-induced electromagnetic sub-cascades, neutrinos, and $\gamma$-rays:
\begin{align*}
                                                \pi^+ \longrightarrow & \mu^+ + \nu_\mu\\
                                                                 & \mu^+ \longrightarrow e^+ + \nu_e + \bar{\nu}_\mu \\
                                                \pi^- \longrightarrow & \mu^- + \bar{\nu}_\mu\\
                                                             & \mu^-  \longrightarrow  e^- + \bar{\nu}_e + \nu_\mu \\
                                                \pi^0  \longrightarrow  &\gamma + \gamma                                              
\end{align*}
These interaction chains provide the connection between cosmic rays, non-thermal $\gamma$-rays, and energetic neutrinos. Non-thermal $\gamma$-rays can not only be produced directly by pion decay of hadronic cosmic rays, but also by radiation mechanisms of cosmic-ray electrons and positrons, which appear as a by-product of hadronic cosmic-ray interactions as secondary component as well as a primary component directly accelerated at the sources. These leptons produce $\gamma$-rays via synchrotron radiation in magnetic fields and inverse Compton scattering off radiation fields.
Typical energies of the secondaries are about 10\% of the nucleon energy for the $\gamma$-rays from pion decay and 5\% for the neutrinos \cite{Ahlers}. While neutrinos travel mostly unimpeded, $\gamma$-rays can undergo further interactions.
They pair produce in radiation fields and thereby initiate electro-magnetic cascades that distribute the original energy to a larger number of 
particles.\\ 
Therefore, photons and neutrinos both trace cosmic-ray interactions. Neutrinos are indicative of hadronic interactions, while $\gamma$-rays 
can be produced by both, hadronic cosmic rays and cosmic-ray electrons. The cumulative spectrum of cosmic rays observed at Earth is the integral of Galactic and extragalactic cosmic rays guided to Earth by propagation mechanisms, the spectrum acting as boundary condition for cosmic-ray physics and to the modelling of cosmic-ray propagation. Neutrinos and $\gamma$-rays on the other hand experience no deflection in magnetic fields due to their neutral nature and thus allow to source the origin of the emission and probe cosmic-ray interactions in situ, enabling the study of particle acceleration mechanisms, the energetics and environments of individual objects.\\ 
With the discovery of gravitational waves in 2015~\cite{GWdetection} a new messenger became available, which is not coupled to the cosmic-ray interactions but rather a thoroughly independent probe of the {\it dynamics} of astrophysical objects. Gravitational waves are associated with the acceleration of massive objects and typical gravitational-wave events are the mergers of compact objects like neutron stars and black holes. The detection of gravitational waves probes 
e.g. the masses of the merging compact objects, their distance, and excentricity, opening a completely new observational window on such coalescence events by providing information on the object type and characteristics.\\\\
The interplay of these messengers happens on all spatial scales, from solar, over Galactic to extragalactic, and similarly on all energy scales, and on all scales the combined information has the potential to reveal insights that are not accessible by the individual observations. However, observational constraints limit the usability of individual messengers. Gravitational-wave measurements are currently limited to frequencies of $\sim 100$~Hz, corresponding to merging binary neutron stars and stellar mass black holes.
For neutrinos, due to the large and irreducible background of atmospheric neutrinos, low-energy neutrinos have a large probability of
being background events and only neutrinos well above 100~TeV can be clearly associated with astrophysical events.
$\gamma$-rays on the other hand are increasingly absorbed when moving to higher energies due to pair production on radiation fields (for the highest energies the cosmic microwave background), limiting the horizon of observable objects.\\
Currently, the multi-messenger approach exploits its full possibilities on the extragalactic scale and the highest energies, where gravitational waves and neutrinos with no significant absorption collect statistics. Particularly exciting results can be expected for the rare cases of such events being close enough to provide also measurements in the full electromagnetic spectrum.\\
In the following I will review multi-messenger onservations on different scales, going from solar system up to redshifts of $z=0.65$. 
\section{Multi-Messenger Searches in Galactic Objects}
On small scales, multi-messenger physics is closely related to the observed cosmic-ray spectrum. 
Already in the 1950s a correlation was observed between solar activity observed in the optical and the flux of cosmic rays.
When the first neutrino observations became feasible the Sun was again the first object to be detected, already probing the potential of multi-messenger observations by tracing not only the Sun's surface or atmosphere in photon observations but also the hydrogen burning in its dense core.
The second source of low-energy neutrinos to be observed was due to an exceptional celestial event, the supernova SN1987A in the Large Magallanic Cloud, which could be detected in optical photons and neutrinos \cite{SN1987A}.\\
In the search for the sources of Galactic cosmic rays, assumed to dominate the local cosmic-ray spectrum at least up to the knee at around a PeV, $\gamma$-ray observations have been used to explore the cosmic-ray origin. Supernova remnants have long been suspected to be prime candidates because of matching energetics 
and with diffusive shock acceleration an acceleration mechanism being in place.
After the first resolved image of a supernova remnant in very-high-energy (VHE) $\gamma$-rays revealed bright emission from the shock front \cite{RXJ_HESS}, the full coverage of the spectral feature of the pion decay of two supernova remnants, W44 and IC443 by Fermi-LAT provided first evidence that supernova remnants accelerate hadronic cosmic rays \cite{PionBump}. The hope for a confirming signal of neutrinos from the direction of SNRs has so far not become reality. Probing the paradigm of a Galactic origin of cosmic rays up to the knee in the CR spectrum in $\gamma$-rays has turned out to be surprisingly difficult. The first ``PeVatron'' has been seen indirectly in the diffuse emission in the Galactic Center region. Cosmic rays that are diffusing out from a source located within the inner 10~pc interact with the dense molecular clouds of that region to produce bright VHE $\gamma$-ray emission. This emission shows a power-law up to 50~TeV with no indication of a cut-off. Solving the transport equation for protons injected at the Galactic Center suggests a pure power-law primary proton spectrum with index of $\sim 2.4$ with a cut-off at 2.9~PeV at 68\% confidence level \cite{PeVatron}.
     
\section{The Sources of High-Energy Neutrinos}
\begin{figure}
\centering
  \includegraphics[width=.9\textwidth]{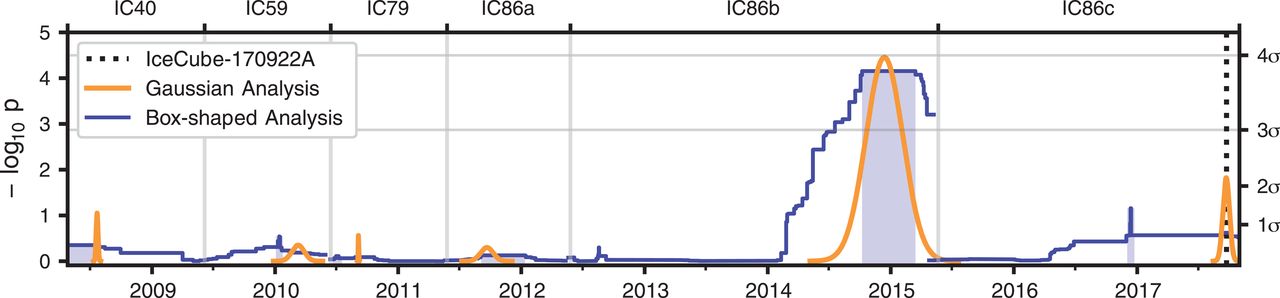}
  \caption{The $p$ value and significance for neutrino outbursts from the direction of TXS 0506+056 as a function of time. Blue and orange correspond to differently shaped time profiles in the analysis. The dotted line indicates the time of IC-170922A. Taken from \cite{IceCubeTXS2}.}
\label{FIG:Neutrino}
\end{figure}

Ever since the announcement of a diffuse flux of astrophysical neutrinos detected by IceCube~\cite{DiffuseNeutrinos} the search has been ongoing for the sources of these neutrinos. The flux is significantly measured at energies above ~20 TeV up to a few PeV~\cite{DiffuseNeutrinos2}.
The isotropic arrival directions point to a predominantly extragalactic origin~\cite{DiffuseNeutrinos2} and no correlation with certain sources is observed \cite{DiffuseNeutrinosCorrelation}.
The picture changed in 2017 with the discovery of a $\sim 290$~TeV muon-track neutrino (IC-170922A) from the direction of the blazar TXS 0506+056, which exhibited a period of enhanced $\gamma$-ray emission at the time of the neutrino, observed by Fermi-LAT. The correlation of the high-energy neutrino and the $\gamma$-ray flare had a significance of $4.1 \sigma$, $3 \sigma$ after trial correction, and provides strong indication that the blazar was responsible for the neutrino event \cite{TXSScience}. The large coverage in electromagnetic observations initiated after this discovery also includes a rise in the TeV emission observed days later by the MAGIC telescopes and allows for a modelling of the combination of the broad-band electromagnetic spectrum and the neutrino. While leptonic scenarios match the electromagnetic radiation, they cannot account for the neutrino emission. Simple hadronic models on the other hand fail to describe the electromagnetic spectrum, especially the observed level of the X-ray emission. Only more complex scenarios fit both \cite{Gao}. This demonstrates that in the combination of messengers a level of detail is obtained that has the potential to probe more realistic scenarios.\\
Motivated by the observed correlation, the IceCube collaboration investigated 9.5 years of archival data at the position of TXS 0506+056 and found a second period of neutrino emission during September 2014 to March 2015, where an estimated $13\pm5$ neutrino candidates were observed \cite{IceCubeTXS2}. The $p$ value of this search as function of time is depicted in Fig.~\ref{FIG:Neutrino}. Only based on the neutrino signal the significance of this earlier outburst is much higher than the IC-170922A event. A chance correlation can be excluded at a confidence level of $3.5 \sigma$. The combination of these two observations provide convincing evidence for TXS 0506+056 being a source of high-energy neutrinos, thereby not only identifying a first source contributing to IceCube's diffuse neutrino flux but also nailing down the first source of extragalactic hadronic cosmic rays and hadronic cosmic rays in general without relying on spectral modelling. Multi-wavelength data on this earlier event is sparse and only data from full-sky instruments continuously running are available. Interestingly, in archival Fermi-LAT data no enhanced $\gamma$-ray emission was seen to accompany the 2014/2015 neutrino outburst. 
This and the observations for IC-170922A pose severe challenges to the modelling of neutrino emission \cite{Rodrigues}.

\section{Multi-Messenger Astronomy with Gravitational Waves}
\begin{figure}
\centering
  \includegraphics[width=.8\textwidth]{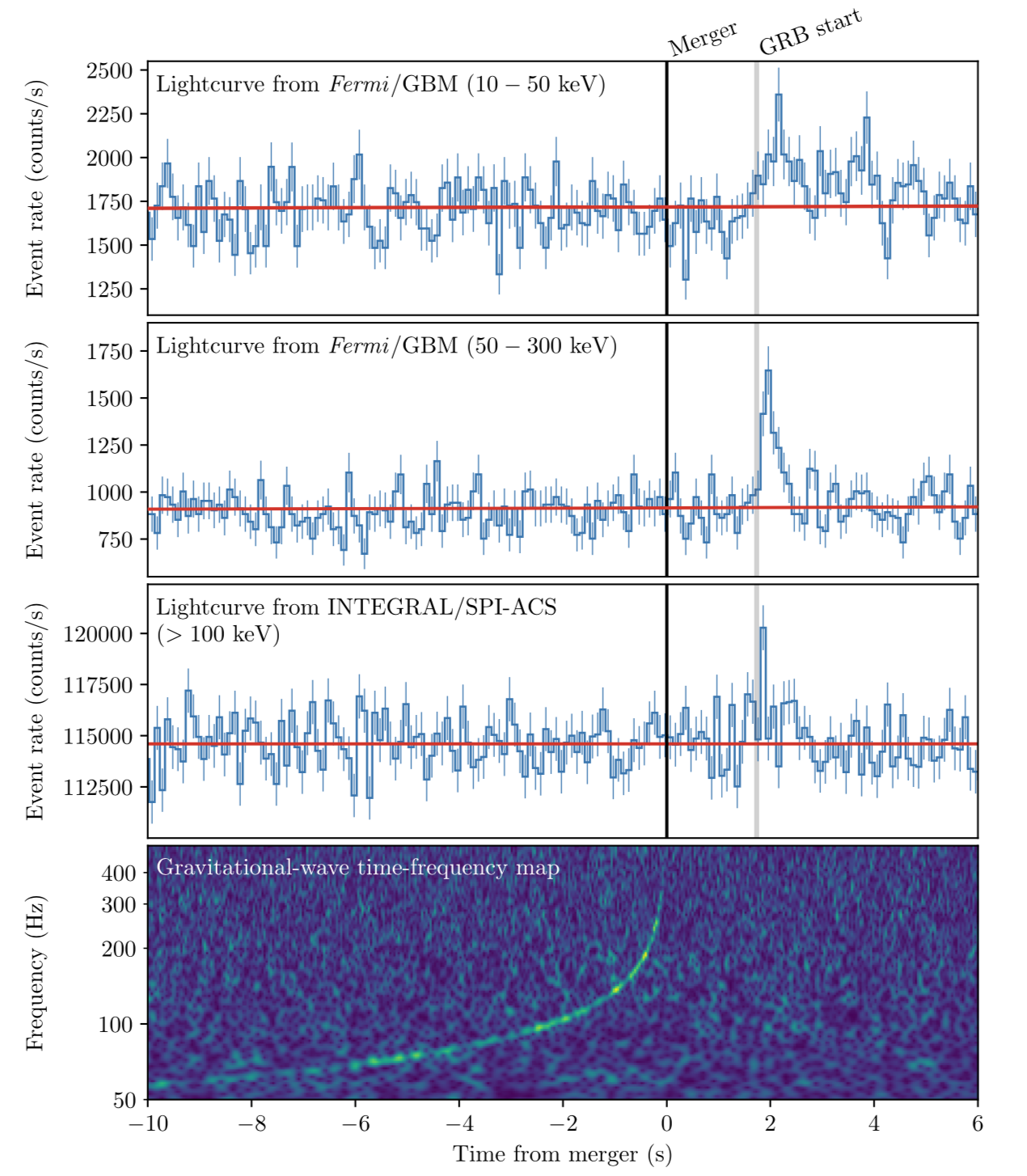}
  \caption{GW170817 in conjunction with the short GRB170817A. Displayed are the light curves of Fermi-GBM at low and high energies (first and second panel), the light curve of Integral (third panel), and the development of the gravitational-wave frequency signal as function of time (bottom panel). Taken from \cite{GW170817}.}

\label{FIG:GW}
\end{figure}

\begin{figure}
\centering
  \includegraphics[width=.7\textwidth]{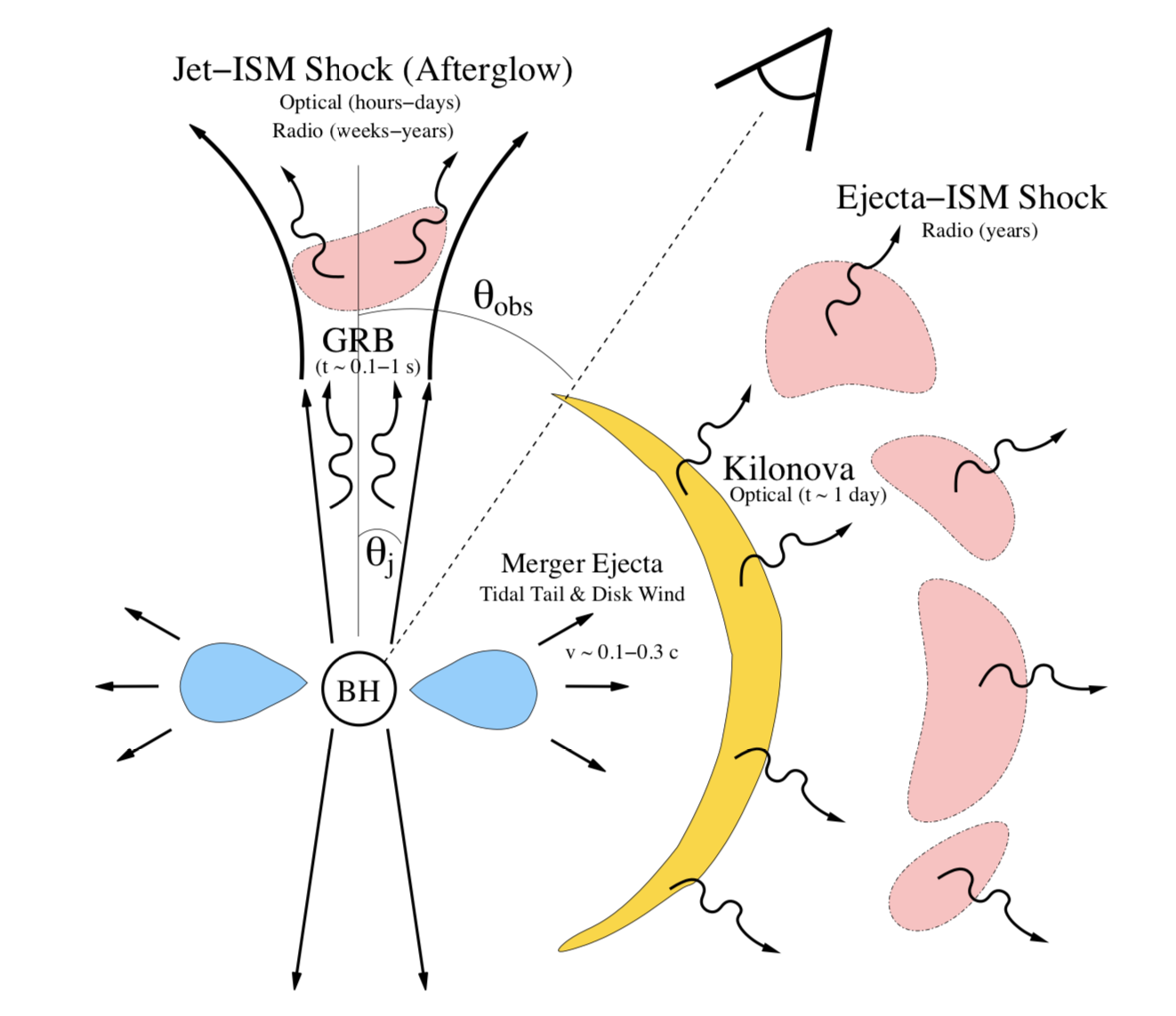}
  \caption{Schematic view of the physical processes taking place in GW170817. Taken from \cite{MetzgerBerger}.}
\label{FIG:Metzger}
\end{figure}

A new era of multi-messenger astronomy started with the gravitational-wave event GW170817, an observation of a neutron star-neutron star merger that turned out to be coincident with a weak short gamma-ray burst (GRB) GRB170817A observed by Fermi-GBM \cite{GW170817}. This event triggered an unprecented follow-up observation campaign throughout the electromagnetic spectrum and beyond with more than 70 instruments joining the campaign \cite{GW170817MM}. In Fig.~\ref{FIG:GW} the gravitational-wave signal is shown together with the light curves of $\gamma$-ray observations of Fermi-GBM and Integral. The results of this campaign are an impressive demonstration of the benefits of the multi-messenger approach. 
Follow-up observations revealed early optical emission discovered within a day of the merger, resulting in the identification of the host galaxy at a distance of approximately 40~Mpc, and, together with infrared emission, the identification of ``kilonova'' features powered by the radioactive decay of r-process nuclei. This provides proof for the long-harboured hypothesis that these elements are synthesized in the neutron-rich environment of neutron-star merger ejecta \cite{MetzgerBerger, Metzger}. A delayed rise of a non-thermal X-ray and radio ``orphan'' GRB afterglow, together with the weakness of the observed short GRB points towards a more powerful off-axis relativistic jet, which was initially beamed away from our line of sight and only with the successive widening of the opening angle included our line of sight in the cone. The schematic diagram of Fig.~\ref{FIG:Metzger} visualises the different contributions from the merger ejecta giving rise to the kilonova and the beam associated with the GRB.
No neutrinos have been detected in the energy range of GeV to EeV as probed by IceCube and ANTARES, as well as above $10^{17}$~eV by AUGER \cite{GW170817MM}.\\
GW170817 has not only confirmed the assumption that binary neutron star mergers produce short GRBs \cite{Bartos}, it has also impressively demonstrated the tremendous potential of multi-messenger astronomy: without the gravitational-wave detection, the Fermi-GBM signal would have been just another weak and poorly localised GRB. Today, it stands out as the closest GRB ever, with the most intensive follow-up data set.


\section{Extending the Electromagnetic Spectrum of GRBs}
While the combination of electromagnetic observations with gravitational waves allowed for the first time to study the progenitors of short GRBs, in 2019 the field of long GRBs was revolutionized with the significant detection of by now three GRBs by ground-based VHE $\gamma$-ray instruments. While ground-based imaging atmospheric Cherenkov telescopes (IACTs) have run extensive GRB follow-up programs for more than a decade (see e.g. \cite{OldGRBPaper, GRBICRC_CH}), only now optimised observation strategies have paid off. The first announcement of GRB190114C by MAGIC \cite{MAGICGRB} was followed by the H.E.S.S. announcement of GRB180720B \cite{HESSGRB}, and the latest addition to this is GRB190829A of August 2019 \cite{HESSGRB2}. Interestingly, the three GRBs probe very different time intervals in the afterglow emission. GRB190114C is a prompt follow-up observation with a minimal time delay of $50$~s, observations spanning till 40~min after the burst and an original detection of $\sim 20 \sigma$, which increased in the offline analysis to $> 50 \sigma$, fully exploiting the advantage of IACTs' huge effective detection areas.
GRB180720B on the other hand went off during daytime in Namibia and was only observed after becoming visible to the H.E.S.S. telescopes 10 hours after the burst. The detection of $\sim 5 \sigma$ at such late time in the afterglow was possible due to the extraordinary brightness of this GRB: the energy flux of the X-ray afterglow is the second brightest after the exceptional GRB130427A. This very late detection of GRB180720B is remarkable since it constitutes a very energetic emission deep in the afterglow phase. Indications of late energetic emission are already seen in the second Fermi-LAT GRB catalog \cite{FERMIGRBCatalog}. 
The latest GRB190829A ranges in between the two with a time delay of four hours and a detection well above $5 \sigma$, but is unusual with respect to its closeness of $z = 0.0785$.
These events span the parameter space of TeV GRB observations with good chances for detection (1) in fast follow-up observations, (2) for very bright objects, and (3) for very close objects.\\
Probing GRBs as extreme accelerators at TeV energies has become reality, even at afterglow time scales of days! 
\section{Technical Challenges of Multi-Messenger Observations}
The basis for multi-messenger observations like the ones described before is the availability of multi-wavelength data for a certain event. While this is straight forward for all-sky instruments, many telescopes have small fields of view and perform pointed observations. Time scales of transient events range from seconds (e.g. the short GRB of GW170817) to days and more (e.g. TXS 0506+056), so in order to catch short-duration events a highly efficient coordination between the instruments, automated reactions, and fast repointing are required. For this purpose, wide field-of-view instruments have set up pipelines to identify potentially interesting events and distribute them in world-wide networks of instruments. In many cases the duty cycle of the instrument, due to the requirement of good atmospheric conditions and darkness, restricts observations as seen e.g. in the case of the late observations of GRB180720B.
Already coming with a certain latency of the alert, also the pointing accuracy of the triggering instrument is of paramount importance for an efficient follow-up observation. For gravitational-wave events, the localization accuracy of interferometers at Earth can be as coarse as hundreds of square degrees. In the case of GW170817 was H.E.S.S. the first ground-based pointed instrument to observe the merger position \cite{GW170817HESS}, only five minutes after the three-instrument localisation of LIGO-Hanford, LIGO-Livingston and VIRGO was circulated. This was possible only due to an optimised observation strategy that combines the localisation uncertainty of the gravitational-wave alert with a galaxy catalog and defines an observation pattern that covers the area in question in tiled observations, prioritised according to the combined likelihood of localisation uncertainty and possibility of a counterpart in form of a galaxy \cite{MonicaMoriond}. \\
While these challenges require careful in-advance preparations of multi-messenger observations and coordination between the instruments, the recent developments and break throughs have provided a strong push in this direction for the entire field and just enabled observation campaigns like the one for GW170817.

\section{Conclusion and Outlook}
Multi-messenger astronomy has seen several major break throughs in the past few years: after the identification of sources of Galactic cosmic rays and the discovery of a first PeVatron, accelerating particles up to the knee in the cosmic-ray spectrum, in $\gamma$-rays, discoveries of the past two years have revolutionised the extragalactic sky. With the discovery of TXS 0506+056 as the first source of high-energy neutrinos, blazars have been established to accelerate hadronic cosmic rays. The detection of gravitational waves and first exploitation of its science potential in the form of the multi-messenger observations of the neutron star-neutron star merger GW170817 confirmed not only the hypothesised connection between binary neutron star mergers and short GRBs, but also revealed a kilonova and established binary neutron star mergers as likely the primary source of heavy r-process elements. Finally, with the detection of GRBs by ground-based TeV $\gamma$-ray instruments, a new window is opened on the study of one of the most extreme phenomena in the Universe.\\
This progress holds great promise for the near future. Technical developments are advancing and so is the international collaboration and coordination between instruments. Furthermore, new instruments with exciting capabilities are just around the corner. For neutrinos, KM3NeT and IceCube-Gen2 are being constructed/in the planning. For gravitational waves, LIGO and VIRGO are currently completing their third observation run O3. KAGRA is close to completion to join these observations and LIGO-India is in preparation.
A multitude of instruments will enrich the observational opportunities in the electromagnetic spectrum. The water Cherenkov telescope HAWC at the highest energies is already taking data for a while, eROSITA has been launched, LSST and SKA are in an advanced state. CTA will place a special focus on transient phenomena as part of the CTA Key Science Projects \cite{CTA}. \\\\
All of these circumstances point to the fact that the triumph of multi-messenger astronomy has just begun and that this rapidly growing field holds a bright future.

\end{document}